\documentclass[12pt]{article}
\usepackage{pdfpages}
\pdfoutput=1
\def\ni{\noindent}
\newtheorem{theorem}{Theorem}[section]
\newtheorem{corollary}[theorem]{Corollary}
\newtheorem{lemma}[theorem]{Lemma}

\def\uv{{\bf{u}}}
\def\vv{{\bf{v}}}

\def\nubm{\mathop{\mbox{\boldmath $\nu$}}}

\def\Abf{{\bf A}\,}

\def\Ibf{{\bf I}\,}
\def\Qbf{{\bf Q}\,}
\def\Qbtl{\tilde{\bf Q}\,}
\def\qtl{\tilde{q}\,}
\newcommand{\bep}[0]{\noindent {\bf Proof}.\ }

\title{
Minimum bias multiple taper spectral estimation
\footnote{
The authors  thank D.~J.~Thomson and the referees for useful comments. 
Research funded by the U.S. Department of Energy.
}
}

\author{
Kurt S. Riedel
  and
Alexander Sidorenko
\\ Courant Institute of Mathematical Sciences, New York University
\\ New York, New York 10012-1185
}

\date{EDICS: SP 3.1.1}

\begin{document}

\maketitle

\begin{abstract}

Two families of orthonormal tapers are proposed for
multitaper spectral analysis: minimum bias tapers,
and sinusoidal tapers $\{ \vv^{(k)}\}$, where
$v_n^{(k)}=\sqrt{\frac{2}{N+1}}\sin\frac{\pi kn}{N+1}$,
and $N$ is the number of points. The resulting sinusoidal multitaper spectral
estimate is  $\hat{S}(f)=\frac{1}{2K(N+1)}
\sum_{j=1}^K |y(f+\frac{j}{2N+2}) -y(f-\frac{j}{2N+2})|^2$,
where $y(f)$ is the  Fourier transform of the stationary time series,
$S(f)$ is the spectral density, and $K$ is the number of tapers.
For fixed $j$, the sinusoidal tapers converge to the minimum bias tapers
like $1/N$.
Since the sinusoidal tapers have analytic expressions, no numerical
eigenvalue decomposition is necessary.
Both the minimum bias and sinusoidal tapers
have no additional parameter for the spectral bandwidth.
The bandwidth of the $j$th taper is simply
$\frac{1}{N}$ centered about the frequencies
$\frac{\pm j}{2N+2}$.
Thus the bandwidth of the multitaper spectral estimate can be
adjusted locally by simply adding or deleting tapers.
The band limited spectral concentration,
$\int_{-w}^w |V(f)|^2 df$,
of both the minimum bias and sinusoidal tapers
is very close to the optimal concentration achieved by the Slepian tapers.
In contrast, the Slepian tapers can have the local bias,
$\int_{-1/2}^{1/2} f^2 |V(f)|^2 df$,
much larger than of the minimum bias tapers and the sinusoidal tapers.
\end{abstract}

\section{Introduction}

We consider a stationary time series, $\{ x_n, n=1\ldots N\}$ with a
spectral density, $S(f)$. A common estimator of the spectral density is to
smooth the square of the discrete Fourier transform (DFT) locally:
\begin{equation}\label{I1}
\hat{S}(f)=\frac{1}{(2L+1)N}
\sum_{j=-L}^L |y(f+\frac{j}{N})|^2,
\end{equation}
where $y(f)$ is the  Fourier transform (FT) of the stationary time series:
$y(f) \; \equiv \; \sum_{n=1}^N x_n e^{-i2\pi nf} \;$.
Since (\ref{I1}) is quadratic in the FT, $y(f)$, it is natural to consider
a more general class of quadratic spectral estimators. We examine quadratic
estimators where the underlying self-adjoint matrix has rank $K$, where
$K$ is prescribed. Using the eigenvector representation, the resulting
quadratic spectral estimator can be recast as a weighted sum of $K$ 
orthonormal rank one spectral estimators. This class of spectral
estimators was originally proposed by  Thomson  \cite{T90}
under the  name of multiple taper spectral analysis (MTSA).
We refer the reader to
\cite{B85,MS90,PLV87,PW93,RST94,T82,T90}
for excellent expositions and generalizations of Thomson's theory.

In MTSA, a rank $K$ quadratic spectral estimate is constructed 
by choosing an orthonormal family of tapers/spectral windows
and then averaging the $K$ estimates of the spectral density.
In practice, only the Slepian tapers
(also known as discrete prolate spheroidal sequences \cite{S78})
are routinely used for MTSA.

In the present paper, we propose and analyze
two new orthonormal families of tapers:
minimum bias (MB) tapers and sinusoidal tapers.
The MB tapers minimize the local frequency bias,
$\int f^2 |V(f)|^2 df$, subject to orthonormality constraints,
where $V(f)$ is the DFT of the taper.
For continuous time,
the MB tapers have simple analytic expressions.
The first taper in the family is Papoulis' optimal taper~\cite{P73}.
For discrete time, the MB tapers satisfy a selfadjoint
eigenvalue problem and may be computed numerically.

In the case of discrete time,
we define
the $k$th sinusoidal taper, $\vv^{(k)}$, as
$v_n^{(k)}=\sqrt{\frac{2}{N+1}}\sin\frac{\pi kn}{N+1}$,
where $N$ is the sequence length.
The
sinusoidal tapers are an orthonormal family
that
converge to the MB tapers with rate $1/N$ as $N\rightarrow\infty$.
These results are given in Section~\ref{MBT}.  
Section~\ref{CSL} compares the local bias,
$\int_{-1/2}^{1/2} f^2 |V(f)|^2 df$,
and the spectral concentration,
$\int_{-w}^w |V(f)|^2 df$,
of the MB tapers, the sinusoidal tapers
and the Slepian tapers.

In Section~5, we show that 
the quadratic spectral estimator which minimizes the expected square local 
error 
is weighted multitaper estimate using the
MB tapers. A local error analysis is given and the optimal number of 
tapers is determined. At frequencies where the spectral density is changing 
rapidly, fewer tapers should be used. 
In Section~\ref{KS}, we  show that kernel smoother spectral estimates 
\cite{GR57,P58} are multitaper estimates and we show that smoothing
the logarithm of the multitaper estimate significantly reduces the
variance in comparison with smoothing athe logarithm of a 
single  taper estimate. We also describe our data adaptive method for 
estimating the spectrum.
In Section~\ref{DATA}, we apply
our spectral estimation techniques to real data and show that
our tapers outperform the Slepian tapers whenever a variable bandwidth
is needed.
In the Appendix, we show that the leading principal components of
kernel smoother spectral estimates  resemble  the MB tapers.


\section{Quadratic Estimators of the Power Spectrum}

Let $N$ discrete measurements, $x_1,x_2,\ldots ,x_N$,
be given as a realization of a stationary stochastic process.
We normalize the time interval between measurements to unity.
The Cramer representation of a discrete stationary stochastic process
\cite{GR57,PW93}
is
\[ x(t) \; = \; \int_{-1/2}^{1/2} e^{2\pi inf} dZ(f) \; , \]
where $dZ$ has independent spectral increments:
$ E[dZ(f)d\overline{Z}(g)] = S(f) \delta (f-g) df dg $.
We assume that the spectral density, $S(f)$, is twice continuously
differentiable.

The spectral inverse problem is to estimate the spectral density,
$S(f)$, given $\{ x_n\}$.
As shown in \cite{B85,MS90},
every quadratic, modulation-invariant power spectrum estimator
has the form:
\begin{equation}\label{E4}
 \widehat{S}(f) \; = \;
   \sum_{n,m=1}^N q_{nm} e^{2\pi i (m-n)f} x_n x_m \; ,
\end{equation}
where $\Qbf=[q_{nm}]$ is a symmetric matrix of order $N$
and does not depend on frequency.
Consider the eigenvector decomposition:
$ \Qbf = \sum_{k=1}^K \mu_k \vv^{(k)}\left(\vv^{(k)}\right)^T $,
where $K$ is the rank of $\Qbf$,
and $\vv^{(1)},\vv^{(2)},\ldots ,\vv^{(K)}$
is an orthogonal system of eigenvectors.
The multitaper representation of the quadratic spectral estimator is
\begin{equation}\label{E6}
 \widehat{S}(f) \; = \;
   \sum_{k=1}^K \mu_k
     \left|\sum_{n=1}^N v_n^{(k)} x_n e^{-2\pi inf}\right|^2 \; .
\end{equation}
In the case $K=1$, estimator (\ref{E6}) turns out
a {\em tapered periodogram estimator}:
\begin{equation}\label{E2}
 \widehat{S}_v(f) \; = \;
 \left|\sum_{n=1}^N v_n x_n e^{-2\pi inf}\right|^2
 \; ,
\end{equation}
with a taper $\vv=(v_1,v_2,\ldots ,v_N)^T$.
If the tapering is uniform (i.e. $v_1=v_2=\ldots =v_N=\frac{1}{\sqrt{N}}$),
we name (\ref{E2}) the {\em periodogram estimator}.
The estimator (\ref{E6}) is a linear
combination of $K$ orthogonal tapered periodogram estimators.
In MTSA, $K$ is normally chosen to be much less than $N$.
The multiple taper
spectral estimate can be thought of as a low rank,
``principal components" approximation of a general quadratic estimator.
Multiple taper analysis has also been applied to nonstationary
spectral analysis \cite{R93,Am94}.

In practice, one does not begin the analysis with a given quadratic estimator,
$\Qbf$. Instead, one usually
{\it specifies a family of orthonormal tapers
$\{ \vv^{(1)},\ldots ,\vv^{(K)} \}$
with desirable properties.}
Previously, only the family of Slepian tapers were used in practice.
The goal of this article is to introduce other families of tapers.

We define the $k$th spectral window, $V^{(k)}$,
to be the FT 
of the $k$th taper:
\begin{equation}\label{E40}
 V^{(k)}(f) \; = \; \sum_{n=1}^N v_n^{(k)} e^{-i2\pi nf} \; .
\end{equation}
The tapers are normally chosen to have their spectral density
localized near zero frequency.
We define two common measures of frequency localization.

The {\em local bias} of a spectral window $V$ is
$\int_{-1/2}^{1/2} f^2 |V(f)|^2 df$.
The term ``local bias'' is used because it is proportional to
the leading order term in the bias error of a taper estimate
as $N\rightarrow\infty$.

The {\em spectral concentration} in band $[-w,w]$ is defined as
$\int_{-w}^w |V(f)|^2 df$.
The bandwidth, $w$, is a free parameter.
The Slepian tapers
are the unique sequences which maximize the spectral concentration
subject to the constraint that they form an orthonormal family.
Detailed analysis of the Slepian sequences is given in \cite{S78}.
We stress that the Slepian tapers depend on the bandwidth
parameter, $w$,
and that the first $2Nw$ spectral windows are concentrated
in the band $[-w,w]$ while the remaining windows
are concentrated outside.


\section{Minimum Bias Tapers}\label{MBT}

\subsection{Continuous Time Case}

We consider time-limited signals;
the time interval is normalized to $[0,1]$.
In the time domain, the taper $\nu (t)$ is a function in
${\cal L}_2[0,1]$ which we normalize to $\int_0^1 \nu^2(t) dt =1$.
The functions $\{\sin (\pi kt),\; k=1,2,\ldots \}$ form
a complete orthogonal basis on $[0,1]$.
(Completeness can be proven by 
extending $\nu(t)$ to be an odd function on $[-1,1]$ and using 
the completeness of the complex exponentials on $[-1,1]$. See   \cite{KF}.)

Setting $a_k = 2 \int_0^1 \nu (t) \sin (\pi kt) dt$,
then $\sum_{k=1}^{K} a_k \sin (\pi kt) dt$ converges to $\nu (t)$
in ${\cal L}_2[0,1]$ as $K\rightarrow\infty$.
The taper normalization is equivalent to
$ \frac{1}{2} \sum_{k=1}^\infty a_k^2 = 1 $.
The Fourier transform of the taper
is the complex-valued, spectral window function:
\begin{equation}\label{E8}
 V(f) \; = \; \int_0^1 \nu (t) e^{-i2\pi ft} dt \; .
\end{equation}
$V(f)$ is defined on the frequency domain $[-\infty ,\infty ]$,
belongs to ${\cal L}_2[-\infty ,\infty ]$,
and satisfies
$ \int_{-\infty}^\infty |V(f)|^2 df =
  2\pi \int_0^1 \nu^2(t) dt = 2\pi $.

The local bias of a taper spectral estimate \cite{GR57,P73,P58} is
\begin{equation}\label{E20}
 E[\widehat{S}(f)] - S(f)
      \; = \;
 \int_{-\infty}^\infty |V(g-f)|^2 (S(g)-S(f)) dg
     \; \approx \;
 \frac{S''(f)}{2} \int_{-\infty}^\infty |V(h)|^2 h^2 dh \; .
\end{equation}
We consider tapers which minimize the leading order term:
\begin{eqnarray}\label{E30}
 \int_{-\infty}^\infty |V(f)|^2 f^2 df
      & = &
 \int_0^1 \left( \frac{1}{2\pi} \frac{d}{dt}
             \sum_{k=1}^\infty a_k \sin (\pi kt) \right)^2 dt
       \nonumber \\
      & = &
 \frac{1}{4} \int_0^1
  \left( \sum_{k=1}^\infty a_k k \sin (\pi kt) \right)^2 dt
   \; = \; \frac{1}{8} \sum_{k=1}^\infty a_k^2 k^2 \; .
\end{eqnarray}
The last expression attains the global minimum when
$a_1^2=2,\; a_2=a_3=\ldots =0$.
Hence, the leading order term in the bias expression (\ref{E20})
is minimal for the taper $\sqrt{2} \sin (\pi t)$
(This result was obtained by Papoulis~\cite{P73}).
In \cite{P73}, Papoulis extends $\nu(t)$ to be zero outside of
$(0,1)$, and therefore has a Fourier integral representation of $\nu(t)$.
We have extended $\nu(t)$ to be periodic and vanish at each integer value.
Since $\nu(t)$ is optimized for  $t \in [0,1]$, both representations are
valid.

Equation (\ref{E30}) implies the more general result: 

\begin{theorem}\label{T1}
$v_k(t)=\sqrt{2}\sin (\pi kt)$ $(k=1,2,\ldots )$
is the only system of functions in ${\cal L}_2[0,1]$
which satisfy the requirements:
\begin{quote}
(i) $\;\;\;\int_0^1 v_k^2(t) dt = 1$, and

(ii) $\;\;\;v_k$ minimizes $\int_{-\infty}^\infty |V^{(k)}(f)|^2 f^2 df$
in the subspace of functions orthogonal to $\;\;\;\;v_1,\ldots ,v_{k-1}$.
\end{quote}
The kth minimum value is
$
 \int_{-\infty}^\infty |V^{(k)}(f)|^2 f^2 df \; = \; \frac{k^2}{4} \; .
$
\end{theorem}

We name
$v_k(t)=\sqrt{2}\sin (\pi kt)$ $(k=1,2,\ldots )$
the {\em continuous time minimum bias tapers}.
The Fourier transform of $v_k(t)$ is
\begin{eqnarray*}
 V^{(k)}(f)
  & = &
   \frac{e^{-i\pi\left( f-\frac{k}{2}\right) }}{i\sqrt{2}} \left\{
       \frac{\sin\left[ \pi\left( f-\frac{k}{2}\right)\right] }
                       {\pi\left( f-\frac{k}{2}\right)}
      - (-1)^k
       \frac{\sin\left[ \pi\left( f+\frac{k}{2}\right)\right] }
                       {\pi\left( f+\frac{k}{2}\right)}
                      \right\}
 \\
  & = &
e^{-i\pi\left( f-\frac{k-1}{2}\right) } \cdot
 \frac{k^2 \sin\left(\pi f - \frac{\pi k}{2}\right)}
      {4\sqrt{2}\pi\left[ f^2 - \left(\frac{k}{2}\right)^2\right] }
      \; .
\end{eqnarray*}
Thus, $|V^{(k)}(f)|$ decays as $f^{-2}$ for large frequencies.

\subsection{Discrete Time Case}

We now consider the discrete time domain $\{ 1,2,\ldots ,N\}$ with 
the corresponding normalized frequency domain $[-\frac{1}{2},\frac{1}{2}]$.
A taper is a vector, $\nubm = (\nu_1,\ldots ,\nu_N)$, normalized by
$\sum_{n=1}^{N} (\nu_n)^2 = 1$.
By the same argument as in the previous section,
the leading order term of the bias of a taper spectral estimate
is proportional to the local bias,
$\int_{-\frac{1}{2}}^{\frac{1}{2}} |V(f)|^2 f^2 df$,
where the frequency window, $V(f)$, is defined in Eq.\ (\ref{E40}).

\begin{lemma}\label{T2}
For the frequency window of a discrete time taper,
\[ \int_{-\frac{1}{2}}^{\frac{1}{2}} |V(f)|^2 f^2 df \; = \;
                                                \nubm \Abf \nubm\ ^* \; , \]
where $\Abf=[a_{nm}]$ with
\[ a_{nm} \; = \;
   \int_{-\frac{1}{2}}^{\frac{1}{2}} e^{i2\pi (n-m)f} f^2 df \; = \;
   \left\{ \begin{array}{l}
             \frac{1}{12} \;\;\;\; if \;\; n=m \; ; \\
             \frac{(-1)^{n-m}}{2\pi^2(n-m)^2} \;\; if \;\; n\neq m \; .
           \end{array}
   \right. \]
\end{lemma}

\begin{corollary}\label{C3}
The tapers $\nubm^{(1)},\ \nubm^{(2)},\ldots , \nubm^{(N)}$,
defined by the requirements
\begin{quote}
(i) $\;\;\;\sum_{n=1}^{N} (\nu_n^{(k)})^2 = 1$,

(ii) $\;\;\;\nubm^{(k)}$ minimizes
$\int_{-\frac{1}{2}}^{\frac{1}{2}} |V^{(k)}(f)|^2 f^2 df$
in the subspace of vectors orthogonal to
$\;\;\;\;\nubm^{(1)} ,\ldots, \nubm^{(k-1)}$,
\end{quote}
are the eigenvectors of the matrix $\Abf$
sorted in the increasing order of the eigenvalues.
The integral
$\int_{-\frac{1}{2}}^{\frac{1}{2}} |V^{(k)}(f)|^2 f^2 df$
is equal to the kth eigenvalue.
\end{corollary}

We name $\nubm^{(1)},\nubm^{(2)},\ldots, \nubm^{(N)}$
the {\em discrete minimum bias tapers}.
They can be approximated by
the {\em sinusoidal tapers}, $\vv^{(1)},\vv^{(2)},\ldots ,\vv^{(N)}$,
which are
discrete analogs of the continuous time minimum bias tapers.
Namely, we define $\vv^{(k)}=(v_1^{(k)},\ldots ,v_N^{(k)})^T$ with
$v_n^{(k)}=\sqrt{\frac{2}{N+1}}\sin\frac{\pi kn}{N+1}$,
$k=1,2,\ldots ,N$.

\begin{lemma}\label{T4}
The sinusoidal tapers, $\vv^{(1)},\vv^{(2)},\ldots ,\vv^{(N)}$,
form an orthonormal basis in ${\bf R}^N$ and the local bias of $\vv^{(k)}$ 
is $\frac{k^2}{4N^2}\left(1 +  {\cal O}(\frac{1}{N})\right)$.
\end{lemma}

\begin{corollary}\label{C5a}
The 
multitaper estimate (\ref{E6}) using $K$ sinusoidal tapers
has local bias equal to 
$\sum_{k=1}^K\mu_k\frac{k^2}{4N^2}+ {\cal O}(\frac{K^2}{N^3})$
and has the following representation:
\begin{eqnarray}\label{E31a}
\hat{S}(f)= 
\sum_{j=1}^K \frac{\mu_j}{2(N+1)} |y(f+\frac{j}{2N+2}) -y(f-\frac{j}{2N+2})|^2 .
\end{eqnarray}
The uniformly weighted estimate, $\mu_k=\frac{1}{K}$,
has local bias  equal to $\frac{K^2}{12N^2}+  {\cal O}(\frac{K^2}{N^3})$.
\end{corollary}

From (\ref{E31a}), the reason for the low bias of  the sinusoidal tapers is 
apparent: {\it the frequency sidelobe from $y(f+\frac{j}{2N+2})$ 
cancels the sidelobe of $y(f-\frac{j}{2N+2})$.} As a result,
the sidelobe of $y(f+\frac{j}{2N+2})$ 
minus $y(f-\frac{j}{2N+2})$ is much smaller than that of the periodogram.

Our preferred weighting is the parabolic weighting: $\mu_j = C (1 - j^2/K^2)$
because the parabolic weighting minimizes the expected square error
in kernel smoothers as $K$ and $N$ tend to infinity. 
Since the weights decrease smoothly to zero, the resulting estimate is 
smooth in frequency.

\begin{corollary}\label{C5b}
The uniformly weighted multitaper estimate using $K$ sinusoidal tapers
can be computed in ${\cal O}(N\ln N) + {\cal O}(KN)$ operations while
the generic multitaper estimate requires  ${\cal O}(KN\ln N)$ operations
plus the cost of computing the $K$ tapers.
\end{corollary}

The following result demonstrates that
the $k$th spectral window is concentrated on
$\left[\frac{k-1}{2(N+1)},\frac{k+1}{2(N+1)}\right] \cup
 \left[-\frac{k+1}{2(N+1)},-\frac{k-1}{2(N+1)}\right]$.

\begin{corollary}\label{T7}
The Fourier transform of $\vv^{(k)}$ equals
\begin{eqnarray*}
 V^{(k)}(f)
  & = &
 \frac{e^{-i\pi \left( (N+1)f-\frac{k}{2}\right)}}{i\sqrt{2(N+1)}}
\left\{\frac{\sin\left[ N\pi\left(f-\frac{k}{2(N+1)}\right)\right]}
            {\sin\left[  \pi\left(f-\frac{k}{2(N+1)}\right)\right]}
       - (-1)^k
       \frac{\sin\left[ N\pi\left(f+\frac{k}{2(N+1)}\right)\right]}
            {\sin\left[  \pi\left(f+\frac{k}{2(N+1)}\right)\right]}
 \right\}
\\
\\
  & = &
 e^{-i\pi \left( (N+1)f-\frac{k-1}{2}\right) } \cdot
  \frac{\sin\frac{\pi k}{N+1}}{\sqrt{2(N+1)}} \cdot
  \frac{\sin\left[ (N+1)\pi f - \frac{\pi k}{2}\right]}
       {\sin^2 (\pi f) - \sin^2 \frac{\pi k}{2(N+1)}}
     \; .
\end{eqnarray*}
\end{corollary}

Thus $|V^{(k)}(f)|=\sqrt{\frac{N+1}{2}}$ for $|f|=\frac{k}{2(N+1)}$,
and $|V^{(k)}(f)|\sim\frac{1}{N^{3/2}}$ for $f={\cal O}(1)$.
In particular, $|V^{(k)}(f)|\approx\frac{\pi k}{\sqrt{2} N^{3/2}}$
when $f\rightarrow\frac{1}{2}$.
In the intermediate frequencies, $\frac{k}{2(N+1)}<f\ll\frac{1}{2}$,
$|V^{(k)}(f)|$ decays as $\frac{1}{f^2}$.

Numerical evaluation (see Table~1) shows that
\mbox{$\|\vv^{(k)}-\nubm^{(k)}\|_{L_2} < \frac{k}{4(N+2)}$}
for all $k$.
The same rate of convergence is observed in $L_{\infty}$ norm:
$\left\| \frac{\vv^{(k)}}{\| \vv^{(k)}\|_{L_{\infty}}}
    - \frac{\nubm^{(k)}}{\|\nubm^{(k)}\|_{L_{\infty}}} \right\|_{L_{\infty}}
     < \frac{k}{2(N+2)}$.
($\|\cdot\|_{L_{\infty}}$ is the supremum norm in the time domain.)
Figure~1 plots the envelopes of $|V^{(k)}(f)|^2$
for both the minimum bias and sinusoidal tapers
with $N=200,\; k=1$.
The spectral energy of both tapers is nearly identical for $|f|<.25$.
Near the Nyquist frequency, the spectral energy of the sinusoidal
taper is roughly three times larger than that of the minimum bias taper.
In the time domain, the sinusoidal tapers are virtually indistinguishable
from the MB tapers.

\section{Comparison of Spectral Localizations}\label{CSL}

We now compare the local bias and the spectral concentration
of three families of orthonormal tapers:
minimum bias (MB) tapers, sinusoidal tapers and Slepian tapers.
Both the local bias and the spectral concentration of the Slepian tapers
depend on the bandwidth parameter, $w$.
For properties of the Slepian tapers, we refer the reader to
\cite{PW93,RST94,S78,T82,T90}.

Since the MB tapers minimize the local bias,
clearly the sinusoidal tapers and the Slepian tapers have larger
local bias. The only question is whether the difference is large or small.
Table~2 gives the local bias, 
$ \sum_{k=1}^K \int_{-1/2}^{1/2} f^2 |V^{(k)}(f)|^2 df$,
of the three families of tapers for $N=50$.
{\em
The sinusoidal tapers come within 0.2\% of achieving the optimal local bias.
}
In contrast, the local bias of the Slepian tapers can be many times larger.

We compute the local bias for three different values of the bandwidth, $w$.
The general pattern is that the $k$th Slepian taper has roughly the same
local bias as the MB taper does when $Nw<k<2Nw$.
The ratio of the local bias of the Slepian tapers to that of the MB tapers
is smallest at $k\approx 1.2Nw$.
As $|k-1.2Nw|$ increases, the local bias rapidly departs
from the optimal value.


Table~3 compares the spectral concentration of the tapers for $N=50$
and $w=.08\:$.
Both the MB tapers and the sinusoidal tapers are within 1.7\%
of the optimal value, except for $k=8,9$. Notice that $2Nw=8$.
Although the ratio of the spectral concentration,
$\int_{-w}^w |V(f)|^2 df$, for the MB and sinusoidal tapers
to that of the Slepian tapers
is usually very close to one,
the ratio of the spectral energy outside of the frequency band $|f|<w$
can be quite large.
Thus our conclusions depend on using
$\int_{-w}^w |V(f)|^2 df$ and not $1-\int_{-w}^w |V(f)|^2 df$
as the measure of frequency concentration.

Figures~1-3 compare $|V^{(k)} (f)|^2$ of the Slepian and MB tapers for
$N=200$. For Figures~1 and~2, we select the Slepian parameter, $w$,
equal to .01 so that $K=2Nw$ equals four. Figure~2 plots $|V_1 (f)|^2$
for the frequencies up to $f=0.14$. The central peak
of the MB taper is more concentrated around $f=0$ than the Slepian
taper is.
The first sidelobe of the MB taper is visible while the first Slepian
sidelobe is much smaller. 

Figure~1 plots the logarithm of $|V_1 (f)|^2$ over the entire frequency
range. The MB taper has smaller range bias in the frequency range $|f|
<0.3w$ and in the frequency range $|f| > 0.13$. In the
middle frequency range, the Slepian taper is clearly better.
The Slepian penalty function maximizes the energy inside
the frequency band, $[-w,w]$,
and thus it is natural that the Slepian tapers do better for $f \sim w$.
By using a discontinuous penalty function,
the Slepian spectral windows experience Gibbs phenomenon and decay only as
$\frac{1}{f},\; (|V(f)|^2\sim \frac{1}{f^2})$.
The MB spectral windows decay as ${1 \over f^2}$, and thus,
it is natural that the MB tapers have lower bias for $f \sim {\cal O}(1)$.

Figure~3 plots $\sum_{k=1}^3 |V^{(k)} (f)|^2$ for $|f| <0.14$ and on this
scale, the MB tapers are clearly preferable to the Slepian tapers.
For larger frequencies,
the energy of the multitaper estimate, $\sum_{k=1}^K |V^{(k)} (f)|^2$,
is very similar to Fig.~2 on the logarithmic scale provided that
$K<<N$.

In summary,
the sinusoidal tapers perform nearly as well as the MB tapers
while the Slepian tapers have several times larger local bias
(except when $k\approx 1.2Nw$).
For $k \ll 2Nw$, the Slepian tapers have better
broad-band bias protection than the minimum bias tapers do.
For $k \sim 2Nw$, the minimum bias tapers provide both smaller local bias
and better broad-band protection due  to the Gibbs phenomena which the
Slepian tapers experience.

\section{Local Error Analysis and Optimal Multitapering}

We now give a local error analysis of MTSA and determine the optimal number
of tapers.
Our results are the multitaper analog of the local error analysis
of the smoothed periodogram \cite{GR57,P58}.
We assume the time series is a Gaussian processes and 
do not consider frequencies near $f=0$ and $f=1/2$. 
In this case, the variance of
the multitaper estimate is approximately
$
{\rm Variance} [ \hat{S} (f)]
   \approx   S(f)^2 \sum_{k=1}^K \mu_k^2$
due to the orthonormality of the tapers

Asymptotically,
the local bias of the multitaper estimate of Eq.\ (\ref{E6}) is
\begin{eqnarray*}
{\rm Bias} [\hat{S}]
& = &
S(f) \left( \sum_{k=1}^K \mu_k -1 \right) + \frac{1}{2}
S''(f) \sum_{k=1}^K \lambda_k \mu_k
 \; ,
\end{eqnarray*}
where $\lambda_k = \int_{-1/2}^{1/2} f^2 |V^{(k)} (f)|^2 df$.
The second term is the MT generalization of (\ref{E20}).
When $\sum_{k=1}^K \mu_k \ne 1$, the MT estimate has bias even in
white noise. When we  require  $\sum_{k=1}^K \mu_k = 1$,
the local expected loss simplifies:

\begin{theorem}\label{T5.4}
For  a Gaussian process, away from $f=0$ and $f=1/2$, 
the  expected square error of the multitaper spectral estimate (\ref{E6}) 
with   $\sum_{k=1}^K \mu_k = 1$ is asymptotically 
(to leading order in $K/N$) 
\begin{equation}\label{E666}
{\rm Bias}^2 + {\rm Variance} \; \approx \; \left[
\frac{1}{2}S''(f) \sum_{k=1}^K \lambda_k \mu_k \right]^2 +
S(f)^2 \sum_{k=1}^K \mu_k^2
\; .
\end{equation}
\end{theorem}

\begin{theorem}\label{D5a}
The multitaper estimate which minimizes the local loss (\ref{E666}) (with 
$\mu_k \ge0$)
is constructed with the minimum bias tapers.
\end{theorem}

Proof: We order the $\mu_k$ such
that $\mu_1\ge \mu_2 \ge \ldots \ge \mu_K$ and define $\mu_{K+1}=0$.
Since the weights, $\mu_k$ are fixed, we need to minimize
$ \sum_{k=1}^K  \mu_k \uv_k \Abf \uv_k^* $ over  all sets of $K$ 
orthonormal tapers, $\uv_1, \ldots, \uv_K$.  
We split the series in $K$ subseries and minimize each subseries separately:
\begin{eqnarray}\label{PR1}
 \min_{\uv_1, \ldots, \uv_K} \sum_{k=1}^K  \mu_k \uv_k \Abf \uv_k^*
 &=& \min_{\uv_1, \ldots, \uv_K} \sum_{k=1}^K 
(\mu_k -\mu_{k-1}) \left(\sum_{j=1}^k  \uv_j \Abf \uv_j^*\right) 
\nonumber \\
&\ge & \sum_{k=1}^K (\mu_k -\mu_{k-1}) 
\left( \min_{\uv_1^{(k)}, \ldots, \uv_k^{(k)} }  
\sum_{j=1}^k  \uv_j^{(k)} \Abf \uv_j^{(k)*}  \right)
\nonumber \\ & = &
\sum_{k=1}^K (\mu_k -\mu_{k-1}) \left(
\sum_{j=1}^k  \lambda_{A,j} \right) \; = \;
\sum_{k=1}^K \mu_k   \lambda_{A,j} ,
\end{eqnarray} 
where the $\lambda_{A,j}$ are the eigenvalues of $\Abf$, given in 
increasing order.
The $\uv_j^{(k)}$ are subject to orthonormality constraints
that  $\uv_j^{(k)}\cdot\uv_{j'}^{(k)} = \delta_{j,j'}$, but are otherwise
independent and minimized separately. 
In the last line of (\ref{PR1}), we use Fan's Theorem \cite{MO79}:
$\min_{\uv_1^{(k)}, \ldots, \uv_k^{(k)} }  
\sum_{j=1}^k  \uv_j^{(k)} \Abf \uv_j^{(k)*}  
= \sum_{j=1}^k  \lambda_{A,j}$, where  the $ \uv_j^{(k)}$ are again subject to
orthonormality constraints.
The theorem is  now proved because the MB tapers are precisely the 
eigenvectors of $\Abf$.

\begin{theorem}\label{D5b}
The uniformly weighted multitaper estimate using $K$ sinusoidal tapers
has an asymptotic local loss of
\begin{equation}\label{E66s}
{\rm Bias^2 + Variance}\ \simeq\ \left[ {S^{\prime\prime} (f)K^2 \over 24N^2}
\right]^2 + {S(f)^2 \over K} 
\; .
\end{equation}
\end{theorem}

\begin{corollary}\label{D5c}
The asymptotic local loss of (\ref{E66s}) is minimized when the
number of tapers is chosen as
\begin{equation}\label{E66K}
K_{opt} \sim \left[ {12 S(f)N^2 \over S^{\prime\prime} (f)}
\right]^{2/5}
\; .
\end{equation}
\end{corollary}

Thus, the optimal number of tapers is proportional to $N^{4/5}$ and varies
with the ratio of $S(f)$ to $S^{\prime\prime}(f)$. Intuitively (\ref{E66K}) 
shows that fewer tapers should be used when the spectrum varies more rapidly.
A key advantage of the MB and sinusoidal tapers is that the tapers need not
be recomputed as $K$ is changed. In contrast, the Slepian tapers are most
efficient when the bandwidth parameter, $w$, is chosen such that
$K \sim 2Nw$. Thus, when the number of tapers is changed, as in (\ref{E66K}), 
the Slepian tapers should be recomputed.

\section{Smoothed Multitaper Estimates}\label{KS}

In our own comparison of kernel smoothing and multitaper estimation
\cite{RST94},
we found that a smoothed multiple taper estimate worked best.
We now evaluate the expected error of the kernel smoothed multitaper
estimator and show that smoothing the logarithm of the multitaper
estimate is useful for estimating the $logarithm$ of the spectrum.

We begin by evaluating that the quadratic estimator  (\ref{E4})
which is equivalent to
a kernel smoother estimates of the spectrum
\cite{GR57,P58}.      
Let $\hat{S}(f)$ be the quadratic spectral estimator (\ref{E4}),
and smooth it with a kernel $\kappa(\cdot)$ of halfwidth $w$ :
\begin{equation}\label{E90a}
\widehat{\widehat{S}}(f) \; = \;
   \int_{-w}^w \kappa({g\over w}) \widehat{S}(f+g)dg \; , 
\end{equation}
where $w$ is the bandwidth parameter and $\kappa(\cdot)$ is a kernel smoother
with domain $[-1,1]$.
This can be rewritten as
\begin{equation}\label{E90}
 \widehat{\widehat{S}}(f) \; = \;
   \sum_{n,m=1}^N \qtl_{nm} e^{i2\pi (m-n)f} x_n x_m \; ,
\end{equation}
where
$ \qtl_{nm} =  q_{nm}\hat{\kappa}_{m-n} $ with 
$\hat{\kappa}_{m} =\ \int_{-w}^w \kappa(g/w) e^{2\pi img} dg $.
Thus smoothing replaces the original quadratic estimator
with matrix $[q_{nm}]$ by another quadratic estimator
with matrix $\Qbtl =[\qtl_{nm}]$.
By Theorem 5.2, this hybrid method cannot outperform the pure multitaper
method with minimum bias tapers.

We now show that combining kernel smoothing with multitapering does improve
the estimation of the $logarithm$ of the spectral density, $\theta (f) = \log
[S(f)]$. One standard approach is to kernel smooth the logarithm of the
tapered periodogram. This approach has the disadvantage that $|y (f)|^2$
has a $\chi_2^2$ distribution and $\log [ \chi_2^2 ]$ has a long lower tail of
its distribution. As a result, $\log [| y(f)|^2 ]$ has an appreciable bias
and its variance is inflated by $\pi^2 /6$. A common alternative is to
estimate the spectrum either by kernel smoothing or by multitapering and
then to take logarithms.
This approach has the disadvantage that the
smoothed spectral estimate tends to be more sensitive to nonlocal bias
effects than the corresponding smoothed log-spectral estimate.

To robustify the log-spectral estimate while reducing the variance inflation
from the long tail, we propose the following hybrid estimate: 1) compute the
multitaper estimate using the sinusoidal tapers with $\mu_k = {1 \over K}$
and then 2) smooth $\hat{\theta}_{MT} (f) \equiv \ln [ \hat{S}_{MT} (f)] -
B_K / K$, where $B_K$ is the bias of
$\ln [ \chi_{2K}^2 ]$. ($B_K \equiv\ \psi (K) - \ln K$ where 
$\psi (K)$ is the digamma function). For white
noise, the variance of $\hat{\theta}_{MT} (f)\  =\ \psi^{\prime} (K)
\underline{\sim}\ {1 \over K} + {1 \over 2K^2}$, so the variance enhancement 
from the logarithm tends rapidly to zero. 

In \cite{RS94b}, we show that the asymptotic error for this
scheme is
\begin{equation}\label{E94}
\theta^{\prime\prime} (f)^2 \left[ b_k w^2 + {K^2 \over 24N^2} \right]^2
+ {C_{\kappa} \over Nw} \left( 1+ {1 \over 2K} \right)^2 \ ,
\end{equation}
where $b_{\kappa}$ and  $C_{\kappa}$ are constants which depend on the kernel
shape. In (\ref{E94}), we assume uniformly weighted sinusoidal tapers
are used and $1\ll K \ll Nw$.
In (\ref{E94}), one factor of $(1+{1 \over 2K})$ is the variance enhancement 
from the
logarithmic transformation and one factor of $(1+{1 \over 2K})$ arises in
the variance calculation of (15) with sinusoidal tapers. Optimizing 
(\ref{E94})
with respect to both $w$ and $K$ yield $w \sim N^{-1/5}$ and $K \sim
N^{8/15}$, thus the smoothing halfwidth $w$ is much larger than $K/N$.
The expected error (\ref{E94}) depends weakly on $K$ provided that $1
\ll K \ll Nw$.

For simplicity, we set $K = N^{8/15}$ and optimize (\ref{E94}) with respect to
the halfwidth $w$. The resulting halfwidth depends on $\theta^{\prime\prime}
(f)$: $w_{opt} ( \theta^{\prime\prime } (f))$ with $w_{opt} \sim
| \theta^{\prime\prime} (f)|^{-2/5} N^{-1/5}$. Thus when the
log-spectrum varies rapidly, the halfwidth should be reduced as
$|\theta^{\prime\prime} (f)|^{-2/5}$.

Since $\theta^{\prime\prime} (f)$ is unknown, we consider two stage
estimators which begin by making a preliminary estimate of
$\theta^{\prime\prime} (f)$ prior to estimating $\theta (f)$.
We then insert the estimate $\widehat{\theta^{\prime\prime} (f)}$ into
the expression for $w_{opt}$: $w(f) = w_{opt} ( \widehat{\theta^{\prime\prime}}
(f))$ and use a variable halfwidth kernel smoother with halfwidth
$\hat{w}(f)$ to estimate $\theta (f)$. Multiple stage kernel estimators
are described in \cite{BGH94,MS87,R93,RS94a,RS94b}.
These multiple stage schemes have a convergence rate
of $N^{-4/5}$ and have a relative convergence rate of at least $N^{-2/9}$.
A more detailed description is given in \cite{BGH94,R93,RS94b}.

\vspace{.1in}

\noindent
\section{Application}\label{DATA}

We now compare spectral estimates on an actual data series. 
We use the microwave
scattering data set which is described in \cite{RST94}. The data measures 
turbulent plasma fluctuations in the Tokamak Fusion Test Reactor
at Princeton. The spectrum is dominated by a 1 MHz
peak which is quasicoherent.
The spectral density varies by over five orders of magnitude.

The bias versus variance trade-off of Sec.~5
shows that fewer tapers should be
used near the peak. To make the spectral estimate smooth,
a parabolic weighting of the tapers is used as described in Sec.~3.
To determine how many tapers to use locally,
we use the multiple stage ``plug-in'' method as described in the 
previous section; i.e. 
we determine the number of tapers
using a pre-estimate on the same data. To reduce the fluctuations from the
estimate of the optimal number, we use a longer data segment to determine
the number of tapers at each frequency. We find the optimal number of
sinusoidal tapers is roughly 24 for frequencies in the 200 to 800 kHz range.
Near the 1 MHz peak, as few as 12 tapers are used to minimize the local
bias error. Between 1300 and 2400 kHz, the spectrum is flatter and  we use 
up to 40  tapers. 

The dotted line is the sinusoidal multitaper estimate, and
the solid, more wiggly, curve is the corresponding Slepian estimate 
using 24 tapers
with $w= 60$ kHz. The 1 MHz peak is poorly resolved in the Slepian estimate, 
and the regions of high curvature are artificially flattened.
For $ f \ge 1.5$ MHz, the Slepian estimate is artificially bumpy
due to statistical noise. 
The variable taper number estimate suppresses these
bumps by averaging over a larger frequency halfwidth. 
We have also used a variable  taper number estimate with the Slepian
tapers. Since the Slepian parameter, $w$ was fixed at 100 kHz to allow for 
forty tapers, the artificial broadening was even more exreme.

Comparing with a converged estimate of the
spectrum based on $N = 45,000 $ shows that the sinuosidal taper estimate is
more accurate. Another significant difference is that the Slepian multitaper
estimate requires much more CPU time than the sinusoidal multitaper
estimate.


\section{Conclusion}

We have proposed and analyzed
the minimum bias and the sinusoidal tapers,
$v_n^{(k)}=\sqrt{\frac{2}{N+1}}\sin\frac{\pi kn}{N+1}$,
for multitaper spectral estimation.
The resulting sinusoidal multitaper spectral
estimate is  $\hat{S}(f)=\frac{1}{2K(N+1)}
\sum_{j=1}^K |y(f+\frac{j}{2N+2}) -y(f-\frac{j}{2N+2})|^2$.
The sinusoidal tapers have low bias because 
{\it the frequency sidelobe from $y(f+\frac{j}{2N+2})$ 
cancels the sidelobe of $y(f-\frac{j}{2N+2})$.} 

The minimum bias tapers minimize the local bias,
$\int_{-1/2}^{1/2} f^2 |V^{(k)}b (f)|^2 df$,
and have good broad-band bias protection as well. Asymptotically,
the quadratic spectral estimate which minimizes the expected local square
error is a multiple taper estimate using the minimal bias tapers.

The sinusoidal tapers have a simple analytic form and approximate
the minimum bias tapers to ${\cal O}\left(\frac{1}{N}\right)$.
The $k$th sinusoidal taper has its spectral energy
concentrated in the frequency bands
$\frac{k-1}{2(N+1)}\leq |f|\leq\frac{k+1}{2(N+1)}$.

The minimum bias
and sinusoidal tapers have no auxiliary bandwidth parameter,
and the bandwidth of the spectral estimate is determined
solely by the number of used tapers.
By adaptively adding and deleting tapers,
a multitaper estimate with the optimal convergence properties
of kernel smoothers can be constructed.
In contrast, the Slepian tapers need to be
recomputed with a different bandwidth. Thus the Slepian tapers are only
practical for fixed bandwidth estimation and this is inherently
inefficient.

\vspace{8mm}

\section{Appendix: Multitaper decomposition of kernel estimates}

In Sec.~6, we showed that kernel smoother estimators (\ref{E90a})
have an equivalent multitaper representation (\ref{E6}) 
We now show that the equivalent multitapers  of some popular kernel
smoother estimates of the spectrum strongly resemble the MB/sinusoidal tapers.
In one special case, this corresondence is exact; i.e. the smoothed periodogram
can be exactly decomposed into MB tapers.

\begin{theorem}\label{T8}
Let $w=\frac{1}{2}$ and $\kappa(f)$ be the parabolic kernel,
$\kappa(f)=\frac{3}{2}-6f^2$.
The eigenvectors of the  kernel smoothed periodogram 
are exactly the discrete minimum bias tapers.
\end{theorem}

Proof: The $\Qbtl$ matrix in {\rm (\ref{E90})} can be  calculated explicitly
for this case. We find $\Qbtl=[b_{nm}]=\frac{1}{N}(\frac{3}{2}\Ibf-6\Abf)$
where $\Abf$ is the matrix from Lemma~\ref{T2}.
Thus $\Qbtl$ and $\Abf$ have the same eigenvectors.

To illustrate that this result is typical
even when we apply a taper and smooth over a small band,
we consider a smoothed tapered periodogram with $N=200$.
We use Tukey's split-cosine taper \cite{RST94}
and then smooth the estimate
with a square box kernel
with a halfwidth of $.01$.
We then evaluate the corresponding $\Qbtl$ matrix and compute its eigenvectors.
Figure~5 displays the first 4 eigenvectors.
They are very close to the sinusoids
$\sqrt{\frac{2}{N+1}}\sin\frac{\pi kn}{N+1}$.
Table~4 shows that this spectral estimate is virtually
a $K=4$ multiple taper spectral estimate.
After $k>4$, the eigenvalues decrease sharply,
and these higher eigenvectors contribute very little
to the overall estimate.


\newpage

\begin{center}
{\bf Table Captions:}
\end{center}

\ni
Table~1: Convergence of the sinusoidal tapers to the minimum bias tapers.

\ni
Table~2: Normalized bias term,
$4(N+1)^2 \sum_{k=1}^K\int_{-1/2}^{1/2} f^2 \sum_{k=1}^Kf^2|V^{(k)}(f)|^2 df$,
for $N=50$.


\ni
Table~3: Spectral concentration,
$\int_{-w}^w \sum_{k=1}^K|V^{(k)}(f)|^2 df$,
for $N=50$.


\ni
Table~4: Eigenvectors of the smooth tapered periodogram estimator.

\vspace{20mm}

\begin{center}
{\bf Figure Captions:}
\end{center}

\noindent
Figure~1: Spectral energy of the minimum bias, sinusoidal and Slepian tapers,
\mbox{$N=200, k=1, w=.01$}.

\noindent
Figure~2: Spectral energy of the minimum bias and sinusoidal tapers,
\mbox{$N=200, k=1$}.

\noindent
Figure~3 Spectral energy,  
$\sum_{k=1}^3 \int_{-1/2}^{1/2} f^2 |V^{(k)}(f)|^2 df$, 
of the minimum bias and Slepian tapers,
\mbox{$N=200, K=3, w=.01$}.



\noindent
Figure~4: Estimated spectral density of the plasma fluctuations.
Dashed line is sinusoidal multitaper estimate and solid line is
estimate using Slepian tapers with $w=60$ kHz. Because the Slepian tapers 
have a fixed bandwidth, the corresponding estimate spectral density
at 1 MHz is artificially broadened while being undersmoothed for $f \ge 1.5$
MHz.

\noindent
Figure~6: First eigenvectors of the smooth tapered periodogram estimator.
\includepdf[pages=-,pagecommand={}]{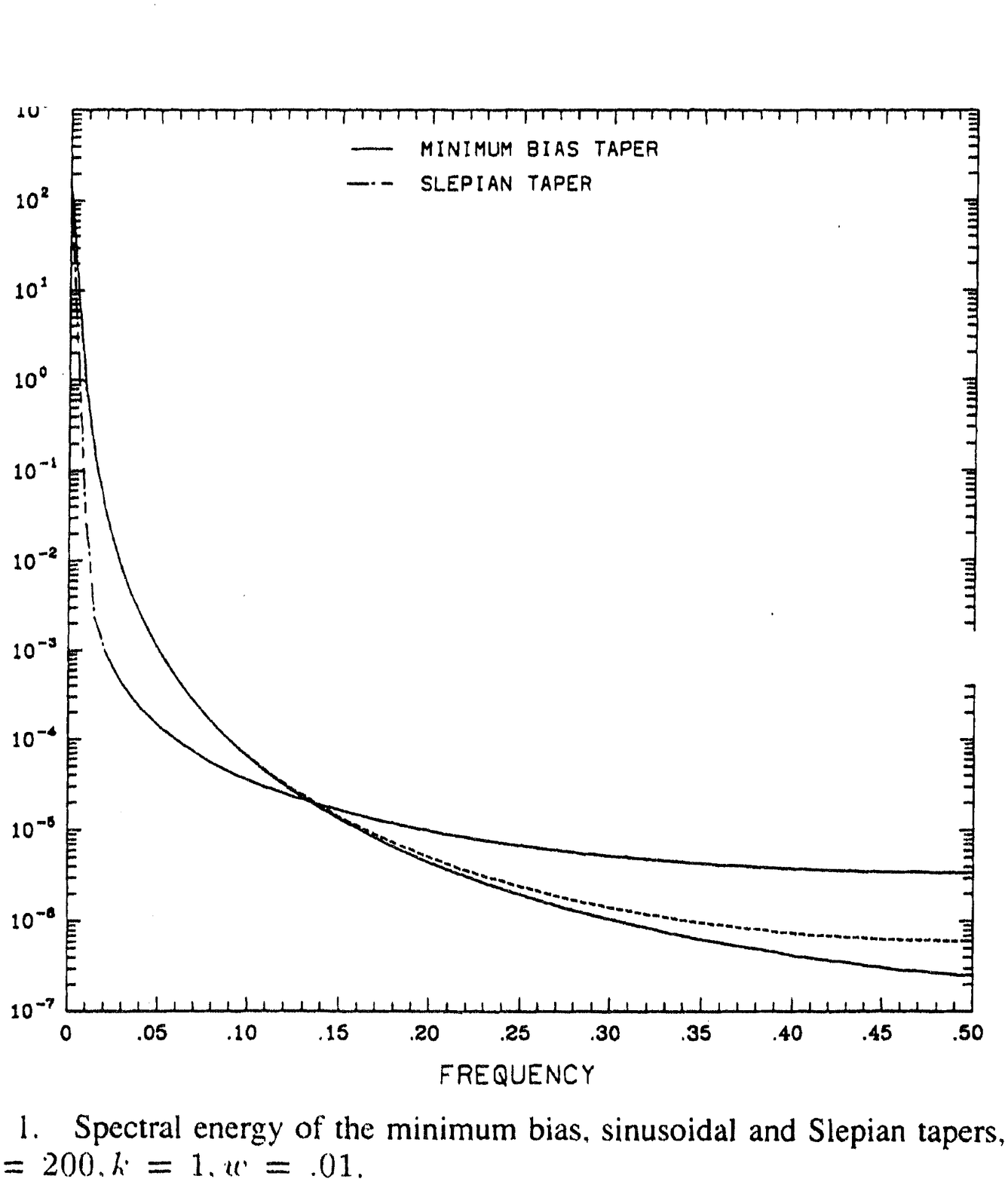}

\newpage

\begin{center}
Table~1. Convergence of the sinusoidal tapers to the minimum bias tapers
\end{center}

\begin{tabular}{|r|c|c|}
\hline
$N$ &
$ \;\;\;\;
 \max_k \left\{ \frac{N+2}{k} \|\vv^{(k)}-\nubm^{(k)}\|_{L_2} \right\}
\;\;\;\; $
&
$ \;\;\;\;
 \max_k \left\{ \frac{N+2}{k}
 \left\| \frac{\vv^{(k)}}{\| \vv^{(k)}\|_{L_{\infty}}}
     - \frac{\nubm^{(k)}}{\|\nubm^{(k)}\|_{L_{\infty}}} \right\|_{L_{\infty}}
    \right\}
\;\;\;\; $
\\ \hline
   20 & 0.24750 & 0.4602
\\  50 & 0.24844 & 0.4760
\\ 200 & 0.24852 & 0.4829
\\ 800 & 0.24844 & 0.4844
\\ \hline
\end{tabular}


\vspace{15mm}

\begin{center}
Table~2. Normalized bias term,
$4(N+1)^2 \sum_{k=1}^K \int_{-1/2}^{1/2} f^2 |V^{(k)}(f)|^2 df$,
for $N=50$
\end{center}

\begin{tabular}{|r|ccccc|}
\hline
$K$ & Minimum & Sinusoidal & & Slepian tapers
&
\\ & bias tapers & tapers
& $w$=0.04
& $w$=0.08 & $w$=0.16
\\
\hline
  1 &    1.0095 &    1.0116 &    1.3439 &    2.6316 &    5.1039
\\
  2 &    5.0475 &    5.0580 &    5.7724 &   10.5484 &   20.4670
\\
  3 &   14.1328 &   14.1622 &   18.0651 &   23.8086 &   46.1953
\\
  4 &   30.2846 &   30.3475 &   58.9520 &   42.5181 &   82.4018
\\
  5 &   55.5217 &   55.6366 &  154.4818 &   66.9996 &  129.2087
\\
  6 &   91.8634 &   92.0528 &  305.4382 &   99.1800 &  186.7507
\\
  7 &  141.3284 &  141.6185 &  496.7959 &  150.0103 &  255.1797
\\
  8 &  205.9362 &  206.3570 &  721.1743 &  251.8833 &  334.6717
\\
  9 &  287.7056 &  288.2899 &  976.5088 &  437.5993 &  425.4379
\\ 
10 &  388.6562 &  389.4409 & 1262.4251 &  702.1523 &  527.7433 \\
\hline
\end{tabular}

\newpage

\begin{center}
Table~3. Spectral concentration,
$\int_{-w}^w |V(f)|^2 df$,
for $N=50,\; w=0.08$
\end{center}

\begin{tabular}{|r|c|c|c|}
\hline
$k$ & Minimum bias tapers& Sinusoidal tapers & Slepian tapers
\\
\hline\hline
 1 & .9997 & .9997 & 1.
\\
 2 & .9988 & .9988 & .9999999
\\
 3 & .9972 & .9972 & .9999989
\\
 4 & .9940 & .9937 & .99997
\\
 5 & .9888 & .9887 & .9995
\\
 6 & .9760 & .9753 & .9928
\\
 7 & .9381 & .9417 & .9380
\\
 8 & .6084 & .6247 & .7002
\\
 9 & .1688 & .1780 & .2981
\\
10 & .0637 & .0624 & .0628 \\
\hline
\end{tabular}

\

\begin{center}
Table~4. Eigenvectors of the smooth tapered periodogram estimator
\end{center}

\begin{tabular}{|c|ccc|}
\hline
$k$ & Weight of the eigenvector & Normalized local bias
& Local bias in comparison with
\\ & $\lambda_k(B)/{\rm tr}(B)$ &
$4(N+1)^2 \int_{-1/2}^{1/2} f^2 |V(f)|^2 df$
& the minimum bias taper (ratio)
\\ \hline \hline
\\  1 &  .2856 &    1.5138 & 1.509
\\  2 &  .2828 &    4.7371 & 1.181
\\  3 &  .2519 &    9.6254 & 1.067
\\  4 &  .1416 &   19.2095 & 1.198
\\  5 &  .0340 &   33.7118 & 1.345
\\  6 &  .0037 &   51.3616 & 1.423
\\  7 &  .0002 &   72.9747 & 1.486 \\
\hline
\end{tabular}

\end{document}